\documentclass[a4paper, 12pt, openany, oneside]{article}
\usepackage[cp1251]{inputenc}
\usepackage[english]{babel}
\usepackage{ graphics,amsfonts, amsmath,bm,amssymb, array, eucal}
\usepackage[dvips]{graphicx}
\usepackage[flushleft,small]{caption2}

\makeatletter
\renewcommand{\@biblabel}[1]{#1.\hfill}
\newcommand{\diag}{\rm \diag\, }
\newcommand{\intl}{\int\limits}

\renewcommand{\Re}{\mathop{\rm Re\,}}

\makeatother {

\begin{document}

\thispagestyle{empty} \large
\renewcommand{\refname}{\begin{center}{\bf REFERENCES}\end{center}}

\begin{center} \bf
{\Large INTERACTION OF THE ELECTROMAGNETIC\\
$ \mathbf{S}$--WAVE WITH THE THIN METAL FILM} \\
\end {center}

\begin{center}
{\it Faculty of Physics and Mathematics,\\ Moscow State Regional
University,  105005,\\ Moscow, Radio st., 10--A}
\end{center}\medskip

\begin{center}
  \bf  A. V. Latyshev\footnote{$avlatyshev@mail.ru$} and
  A. A. Yushkanov\footnote{$yushkanov@inbox.ru$}
\end{center}\medskip

\begin{abstract}
It is shown that for thin metal films, thickness of which does not
exceed a thickness of a skin --  layer, the problem allows
analytical solution for any boundary conditions. The analysis of
transmission, reflection and absorption of an electromagnetic wave
coefficients depending on a angle of incidence, thickness of a
layer, coefficient of specular reflection and frequency of
oscillations of electromagnetic field is carried out.
\medskip

{\bf Key words:} degenerate collisional plasma, electromagnetic
$p$--wave, thin metallic film, coefficients of transmission,
reflection and absorbtion.
\medskip

PACS numbers:  73.50.-h   Electronic transport phenomena in thin
films, 73.50.Mx   High-frequency effects; plasma effects,
73.61.-r   Electrical properties of specific thin films,
73.63.-b   Electronic transport in nanoscale materials and
structures
\end{abstract}

\begin{center}
\bf  INTRODUCTION
\end{center}

The problem of interaction of an electromagnetic wave with the metal
films attracts attention to itself for a long time \cite{F69} --
\cite{Landau8}. It is connected with the theoretical interest to
this problem, and with numerous practical appendices of it as well.

Research of interaction of an electromagnetic wave with conducting
medium (in particular, metal films) was carried out basically for a
case of specular dissipation of electrons on a film surface. It is
connected with the fact that for more general boundary conditions
the problem becomes essentially complicated and does not allow
analytical solution generally. At the same time for the real
materials coefficient of specular electron reflection from the
surface is far from unit as a rule. For example, in the work
\cite{Zav} on the basis of the analysis of longitudinal magnetic
resistance of the thin metal wire it is shown, that for sodium the
coefficient of specular reflection is equal to 0.3.

In the present work it is shown that for thin films, thickness of
which does not exceed thickness of a skin -- layer, the problem
allows analytical solution for any boundary conditions.

Let's notice, that the most part of reasonings carrying out below is
fair for more general case of conducting (in particular,
semi-conductor) films.\medskip

\begin{center}
\bf  PROBLEM STATEMENT
\end{center}

Let's consider a thin slab of metal, which the electromagnetic wave
falls on. An angle of falling  we will designate as $\theta$. Let's
assume that a vector of electric field of the electromagnetic wave
is parallel to a slab surface. Such wave is called $s$--wave
(see \cite{K} or \cite {F69}).

We take the Cartesian system of coordinates with the origin of
coordinates on one of the slab surfaces, an axis $x$, directed deep
into a slab. Axis $y$ we direct parallel to electric field vector
of electromagnetic wave. Then the behaviour of electric and magnetic
fields of a wave in a layer is described by the following system of
differential equations \cite{K}:
$$
\left\{\begin{array}{l}
\dfrac{dE_y}{dx}-ikH_z=0 \\\\
\dfrac{dH_z}{dx} +ik(\sin^2\theta-1)E_y =-\dfrac{4\pi}{c}j_y.
\end{array} \right.
\eqno {(1)}
$$

Here $c$ is the velocity of light, $\,\bf j$ is the current density,
$k$ is the wave number ($k =\dfrac{\omega}{c}$),
$E_y(x)$ and $H_z(x)$ defined from relations for electric and
magnetic fields
$$
\mathbf{E}=e^{-i\omega t+ik\sin \theta y}\{0,E_y(x),0\},\qquad
\mathbf{H}=e^{-i\omega t+ikz}\{H_x(x),0,H_z(x)\}.
$$

Let's designate a thickness of a slab as $d$.

Transmission coefficient $T$, reflection coefficient $R$ and slab
absorption coefficient $A$ of the electromagnetic wave are described
by the following expressions \cite{F69}, \cite{F66}
$$
T=\dfrac{1}{4}\big|P^{(1)}-P^{(2)}\big|^2,
\eqno {(2a)}
$$
$$
R=\dfrac{1}{4}\big|P^{(1)} +P^{(2)}\big|^2,
\eqno {(2b)}
$$
$$
A=1-T-R.
\eqno {(2c)}
$$

Quantities $P^{(j)} \; (j=1,2)$ are defined by the following
expressions
$$
P^{(j)}=\dfrac{Z^{(j)}\cos\theta-1}{Z^{(j)}\cos\theta+1},
\qquad
j=1,2.
\eqno {(3)}
$$

Quantities $Z^{(1)}$ and $Z^{(2)}$ correspond to an impedance on the
bottom slab surface at antisymmetric by electric field (case 1, when
$E_y(0)=-E_y (d), H_z(0)=H_z(d)$) and symmetric by electric field
(case 2, when $E_y(0)=E_y(d), H_z(0)=-H_z(d)$) configurations of
external fields.

The impedance thus in both cases is defined as follows
$$
Z^{(j)}=\dfrac{E_y(-0)}{H_z(-0)}.
\eqno{(4)}
$$

\begin{center}
\bf  TRANSMISSION COEFFICIENT, REFLECTION COEFFICIENT
AND ABSORPTION COEFFICIENT IN THE THIN SLAB
\end{center}

Let's consider a case when the width of a slab $d$ is less than
depth of a skin -- layer $\delta$. We will note, that depth of a
skin -- layer depends essentially on frequency of radiation,
decreasing monotonously during the process of growth of the last.
The quantity $\delta$ possesses the minimum value in so-called
infra-red case \cite{Landau10}
$$
\delta_0 =\dfrac {c} {\omega_p},
$$
where $\omega_p$ is the plasma frequency.

For typical metals \cite{Landau10} $\delta_0\sim 10^{-5}$ cm.

Thus for the films thickness of which $d$ is less than $\delta_0$
our assumption is correct for any frequencies.

Electric and magnetic fields vary little at distances less than
depth of a skin -- layer. Therefore under fulfilment of the given
assumption $d<\delta$ this field will vary a little within a slab.
In case 1 when $H_z(0)=H_z(d)$, it is possible to assume, that
quantity $H_z$ is constant in the slab. Change of quantity of
electric field at the thickness of a slab can be defined from the
first equation of system (1)
$$
E_y(d)-E_y(0)=ikdH_z.
$$

Considering antisymmetric character of electric field in this case
we receive
$$
E_y(0)=-\dfrac{ikdH_z}{2}.
$$

Accordingly (4) for the impedance we have
$$
Z^{(1)}=-\dfrac{idk}{2}.
\eqno{(5)}
$$

For the case 2 when $E_y(0)=E_y(d)$,  it is possible to consider
electric field as constant in the slab, which we will designate as
$E_y$. Then magnetic field change at the width of slab can be
defined from the second equation of the system (1)
$$
H_z(d)-H_z(0)=ikd(\sin^2\theta-1)E_y-\dfrac{4\pi}{c}\intl_0^d
j_y(x)dx.
\eqno{(6)}
$$

Thus
$$
j_y(x)=\sigma(x)E_y,
$$
where $\sigma(x)$ is the conductivity which depends from coordinate $x$ in
general case.

Let's introduce the conductivity averaged by thickness of slab
$$
\sigma_d =\dfrac{1}{E_y d}\intl_0^d j_y(x)dx =
\dfrac{1}{d}\intl_0^d\sigma(x)dx.
\eqno{(7)}
$$

Then the relation (6) according to (7) can be rewritten in the form
$$
H_z(d)-H_z(0)=ikd(\sin^2\theta-1)E_y-\dfrac{4\pi d\sigma_d}{c}E_y.
$$

Considering symmetry of the magnetic field, from here we receive
$$
H_z(0)=-\dfrac{1}{2}ikd(\sin^2\theta-1)E_y +
\dfrac{2\pi d\sigma_d}{c}E_y.
$$

For the impedance (4) we receive
$$
Z^{(2)}=\dfrac{2c}{-ickd(\sin^2\theta-1)+4\pi d\sigma_d}.
$$

Let's assume further, that length of the wave of incoming radiation
surpasses essentially thickness of a slab. This assumption is
satisfied for the majority of cases when the thickness of a slab is
less than the depth of a skin -- layer. Then the quantity $kd\ll 1$
and in expressions for impedances it is possible for to neglect it.
It is hence received according to (5)
$Z^{(1)}=0$, and $Z^{(2)}=\dfrac{c}{2\pi d\sigma_d}$.
According to (3) we have
$$
P^{(1)}=\dfrac{c\cos\theta-2\pi d\sigma_d}
{c\cos\theta+2\pi d\sigma_d}, \qquad P^{(2)}=-1.
$$

According to the expressions (2a), (2b) and (2c) we receive

$$
T=\Big|\dfrac {c\cos\theta}
{c\cos\theta+2\pi \sigma_d d}\Big|^2,
\eqno {(8a)}
$$
$$
R =\Big|\dfrac{2\pi \sigma_d d}{c\cos\theta+2\pi \sigma_d d}
\Big|^2,
\eqno {(8b)}
$$
$$
A =\dfrac{4c\pi \Re(\sigma_d) d\cos\theta}
{\big|c\cos\theta+2\pi \sigma_d d\big|^2}.
\eqno {(8c)}
$$

In limiting case of a non-conductive slab, when $\sigma_d\to 0$ from
these expressions we have $T\to 1, \, R\to 0, \, A\to 0$. At almost
tangential falling, when $\theta\to \pi/2$ we receive $T\to 0, \,
R\to 1, \, A\to 0$.

If we designate
$$
B =\dfrac {2\pi d \sigma_d} {c\cos \theta},
$$
then formulas (8) can be written down in the compact form
$$
T =\dfrac{1}{|1+B |^2}, \qquad R=\dfrac{1}{|1+B^{-1}|^2}, \qquad
A =\dfrac{2\Re B}{|1+B|^2}.
\eqno {(9)}
$$

Let's consider a case of a metal film. Let the relation $kd\ll 1$ to
be satisfied. Then in a low-frequency case, when $\omega\to 0$, the
quantity $\sigma_d$ can be presented in the form \cite{S}
$$
\sigma_d =\dfrac {w}{\Phi(w)} \, \sigma_0, \quad\quad
w=\dfrac{d}{l}.
\eqno {(10)}
$$

Here
$$
\dfrac {1} {\Phi(w)}=\dfrac{1}{w}-\dfrac{3}{2w^2}(1-p)
\intl_1^\infty\Big(
\dfrac {1}{t^3}-\dfrac{1}{t^5}\Big)\dfrac{1-e^{-wt}}{1-pe^{-wt}}dt.
$$

Here $l$ is the mean free electron path,
 $p$ is the reflectivity coefficient, $\sigma_0 =\omega_p^2\tau/(4\pi)$
is the static conductivity of the volume sample, $\tau=l/v_F$ is the
time of the mean free electron path, $v_F$ is the electron speed on
Fermi's surface. It is supposed, that Fermi's surface has spherical
form.

In a low-frequency case when the formula (10) is applicable,
coefficients $T, R, A$ according to the formulas (8) do not depend
on frequency of the incident radiation.

For any frequencies of the expression (8) will be satisfied under
the condition, that it is necessary to use the following expression
as quantity $l$
$
l\to \dfrac{v_F \tau}{1-i\omega\tau},
$
and instead of $\sigma_0$  we should use the expression
$
\sigma_0\to \dfrac{\sigma_0}{1-i\omega\tau}.
$

In case of arbitrary frequencies coefficients of transmission,
reflection and absorption are calculated also with the help of the
formulas (9), in which
$$
w =\dfrac{d}{l}(1-i\omega\tau),
$$
and
$$
B =\dfrac{2\pi d\sigma_0}{c\cos \theta (1-i\omega\tau)}
\Bigg[1-\dfrac {1.5}{w}(1-p)
\int\limits_{1}^{\infty}\Big(\dfrac {1}{t^3}-\dfrac{1}{t^5}\Big)
\dfrac {1-e^{-wt}}{1-pe^{-wt}}dt\Bigg].
$$

Let's consider the case of a thin slab of sodium. Then \cite{F69}
$\omega_p=6.5\cdot 10^{15}$ sec$^{-1}$, $v_F=8.52\cdot 10^7$ cm/sec.
Frequency of the volume collisions of electrons we take to be equal
$\nu=\tau^{-1}=10^{-3}\omega_p$\,sec$^{-1}$ .
On fig. 1, 2 and 3 correspondingly
the dependence of coefficients of transmission, reflection and
absorption on an angle of falling of an electromagnetic wave on the
border, on a thickness of a layer and on reflectivity coefficients
is represented. On fig. 4 and 5 dependence of reflectivity on
frequency of oscillations of an electromagnetic field under various
values of thickness of the slab is represented. On fig. 4 the case
when coefficient of specular reflection is equal to zero is
considered. On fig. 5 the case when coefficient of specular
reflection is equal to unit is considered.

\begin{center}\bf
CONCLUSION
\end{center}

From fig. 1 it is visible, that dependences of all the coefficients
$T=T (\theta) $, $R=R (\theta) $ and $A=A (\theta) $ become apparent
close to $ \theta =\dfrac {\pi} {2} $. Thus the absorption
coefficients has smooth maximum close to the point $ \theta =\dfrac
{\pi} {2} $.

It is interesting to note (fig. 2), that the quantity of absorption
coefficient practically does not depend on a thickness of the slab
$d$ (under change $d$ from $10^{-7}$ cm to $10^{-6}$ cm). Under this
change $d$ approximately in 2 times the propagation coefficient
decreases, and the reflection coefficient increases (also in 2
times).

On fig. 3 for the first time dependence of coefficient
$T=T (p)$, $R=R(p)$ and $A=A(p)$  on quantity of coefficient of specular
reflection is found out.
Coefficients $T, R, A $ discover strong dependence on coefficient of
specular reflection that is found out for the first time. With the
growth of coefficient of specular reflection the reflection
coefficient increases, and the absorption coefficient decreases.

The analysis of graphs on fig. 4 and fig. 5 shows, that with growth
of oscillations frequency the reflection coefficient is monotonously
decreasing function. With growth of a thickness of the slab the
values of reflection coefficient increase, with growth of
coefficient of reflection the values of reflection coefficient also
increase.

\begin{figure}[h] \center
\includegraphics[width=16.0cm, height=8cm]{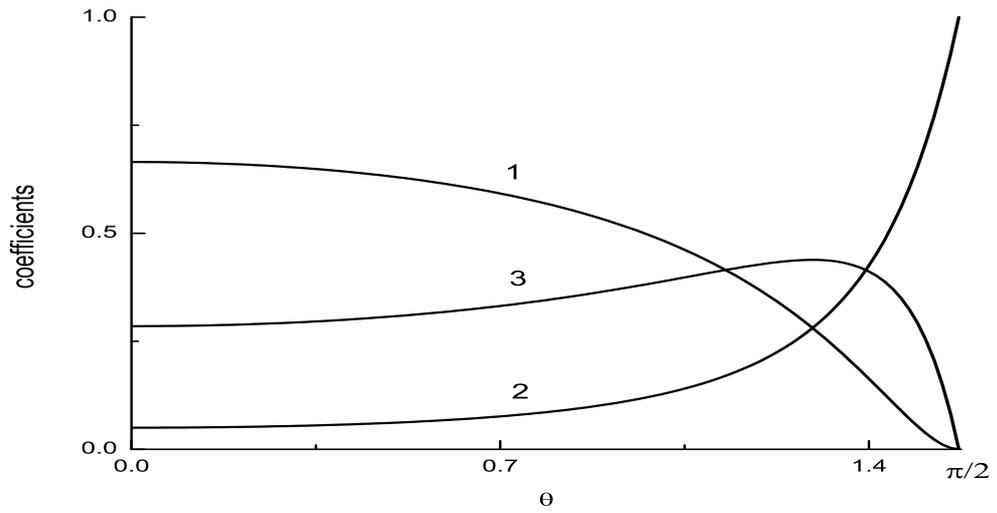}
\caption{Dependence of transmission coefficient (curve
$1$), reflecion coefficient (curve $2$) and absorption coefficient
(curve $3$) on quantity of the angle of incidence
$\theta$, $0\leqslant \theta \leqslant \dfrac{\pi}{2}$, $d=10^{-7}$ cm,
$\omega=10^{-2}\omega_p\, sec^{-1}$, $p=0.5$.}
\label {rateIII}
\end{figure}

\begin{figure} [t]
\begin{center}
\includegraphics[width=16.0cm, height=8cm]{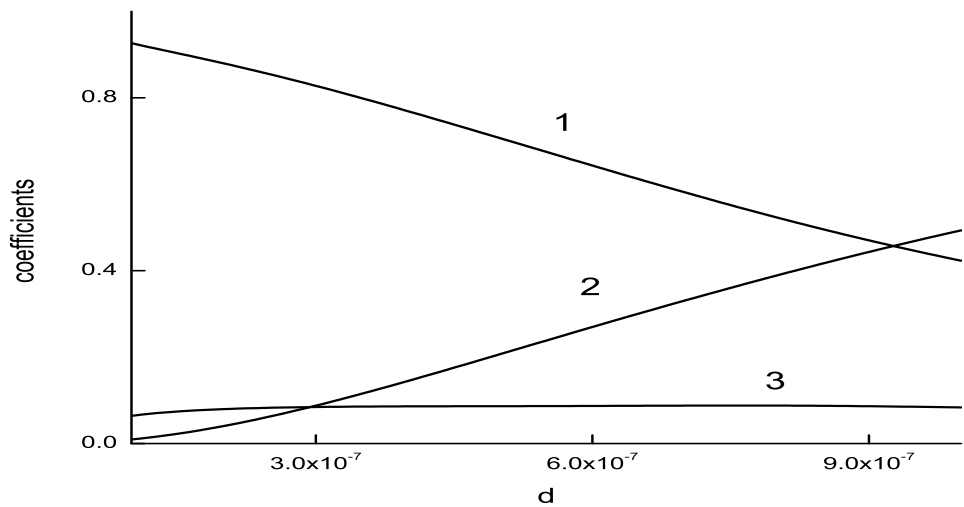}
\caption {Dependence of transmission coefficient (curve
$1$), reflecion coefficient (curve $2$) and absorption coefficient
(curve $3$) on quantity of the thickness of the slab $d$,
$10^{-7}$ cm $\leqslant d \leqslant $ $10^{-6}$ cm,
at normal falling of the wave ($ \theta=0$),
$\omega=10^{-1}\omega_p\, sec^{-1}$, $p=0.5$.}
\label {rateIII}
\end{center}
\end{figure}

\begin{figure}[b]
\begin{center}
\includegraphics [width=16.0cm, height=14cm]{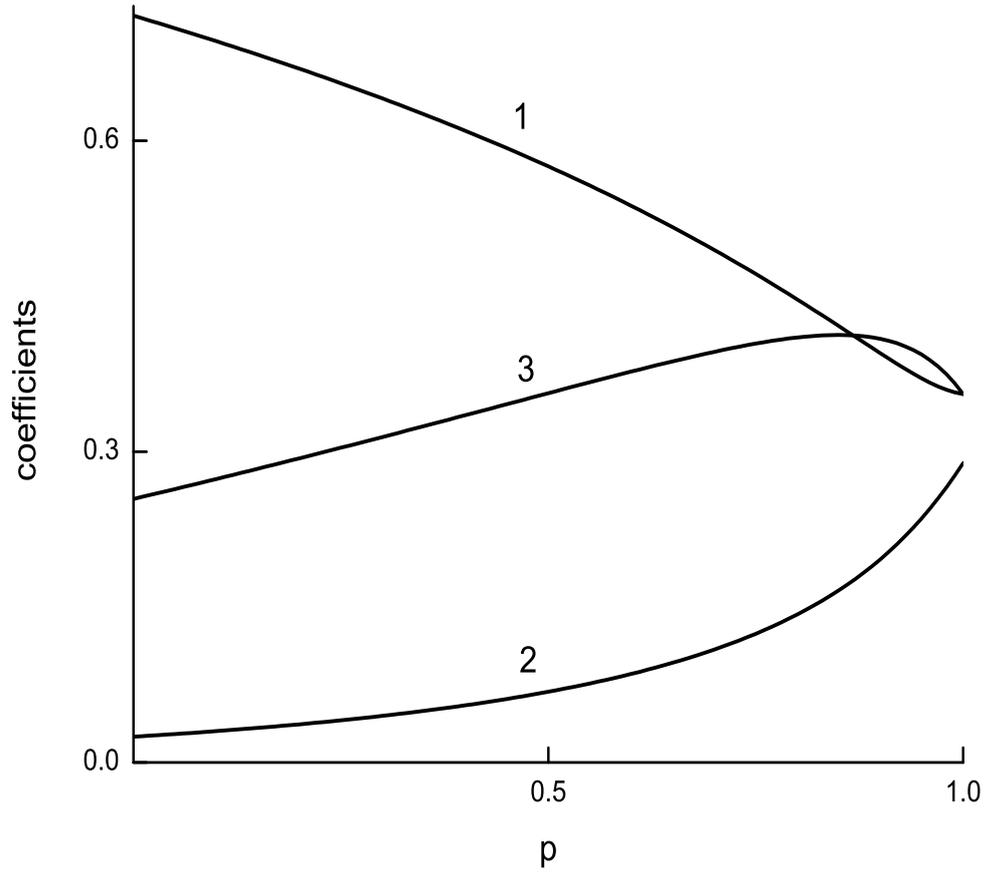}
\caption {Dependence of transmission coefficient (curve
$1$), reflecion coefficient (curve $2$) and absorption coefficient
(curve $3$) on quantity of coefficient of specular reflection
$p$ (0$\leqslant p \leqslant 1$) under normal incidence of the wave
($\theta=0$), $ \omega=10^{-1}\omega_p\, c^{-1}$, $d=10^{-7}$ sm}.
\label {rateIII}
\end{center}
\end{figure}

\begin{figure}[t]
\begin{center}
\includegraphics[width=16.0cm, height=8cm]{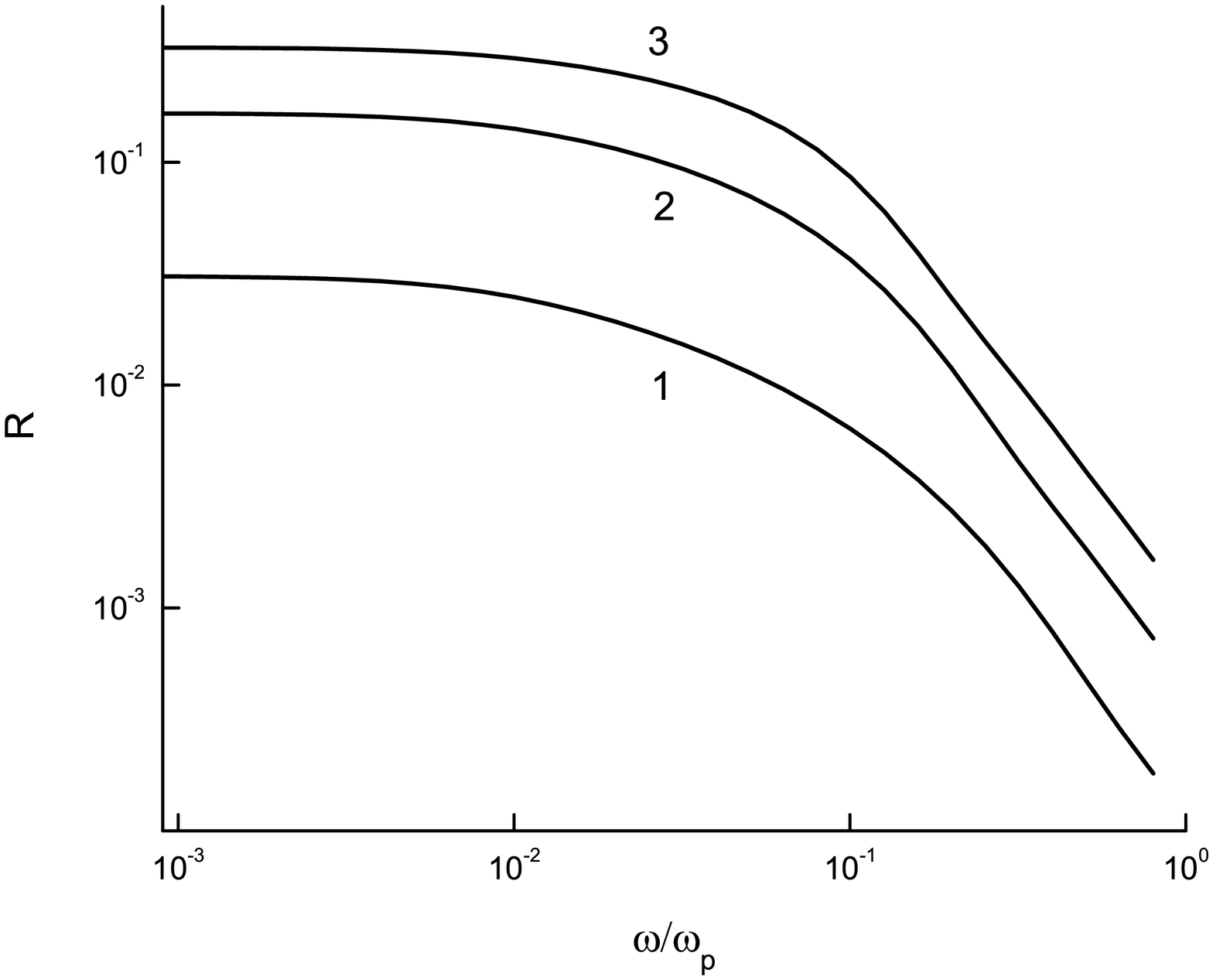}
\caption{Dependence of reflection coefficient $R$ on the
quantity of oscillation frequency of the field $\omega $ under various
values of a thickness
slab $d$ and under normal falling of an electromagnetic wave ($\theta=0$).
Curves of $1,2,3$ correspond to values
$d=10^{-7}$ cm, $2\cdot 10^{-7}$ cm, $3\cdot10^{-7}$ cm
The coefficient of specular reflection is equal to zero ($p=0$).}
\label {rateIII}
\includegraphics[width=16cm, height=8cm]{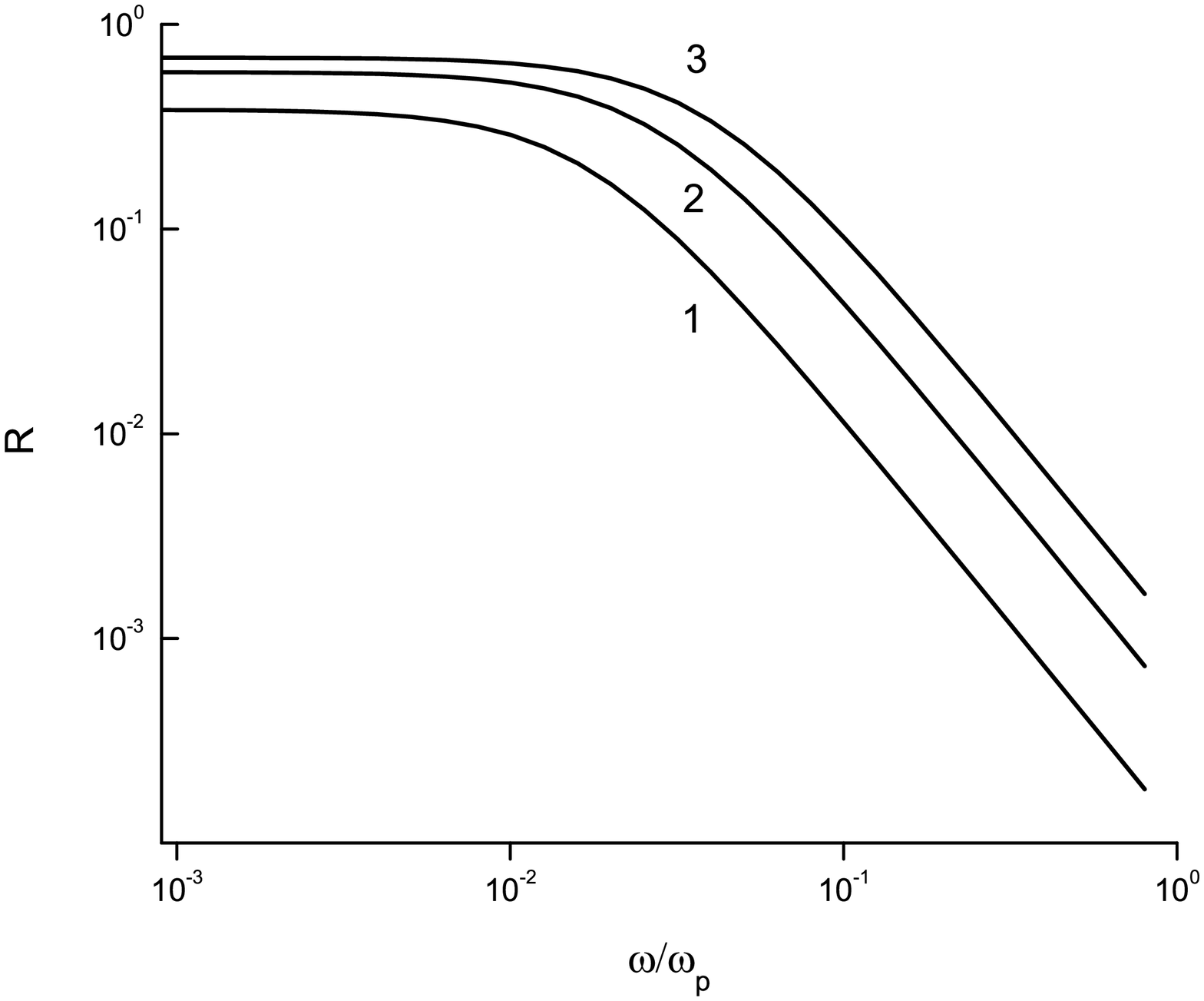}
\caption{Dependence of reflection coefficient $R$ on the quantity of
oscillation frequency of the field $\omega $ under various values of
a thickness slab $d$ and under normal incidence of an
electromagnetic wave ($\theta=0$). Curves of $1,2,3$ correspond to
values $d=10^{-7}$ cm, $2\cdot 10^{-7}$ cm, $3\cdot10^{-7}$ cm. The
coefficient of specular reflection is equal to unit ($p=1$).} \label
{rateIII}
\end{center}
\end{figure}


\clearpage

\begin {thebibliography}{99}

\bibitem {F69} {\it Jones W. E., Kliewer K. L., Fuchs R.}
Nonlocal theory of the optical properties of thin metallic films
//
Phys. Rev. 1969. Vol. 178. No. 3. P. 1201 - 1203.

\bibitem {F69-2} {\it Kliewer K. L., Fuchs R.}
Optical properties of an electron gas: Further studies of a nonlocal
description
//Phys. Rev. 1969. Vol. 185. No. 3. P. 805 - 913.

\bibitem {K} {\it Kondratenko A. N.} Penetration of waves in
Plasma. M: Atomizdat, 1979. 232 P. (in russian).

\bibitem {F09} {\it
Paredes-Ju\'{a}rez A.,  D\'{i}as-Monge F., N. M. Makarov N. M.,
P\'{e}res-Rodr\'{i}gues F.} Nonlocal effects
in the electrodynamics of metallic slabs. JETP Lett, 90:9
(2010), 623--627.

\bibitem {Landau8} {\it Landau L. D., Lifshits E. M.}
Electrodynamics of Continuous Media,
Butterworth-Heinemann (Jan 1984). P. 460.

\bibitem {Zav} {\it Zavitaev E. V., Yushkanov A. A.}
Dependence of the electric conductivity of a thin cylindrical wire in a
longitudinal magnetic field on the character of electron reflection
//Journal of Experimental and Theoretical Physics, Volume 103,
Issue 5, pp.768-774.


\bibitem {F66} {\it Fuchs R., Kliewer K. L., Pardee W. J.}
Optical properties of an ionic crystal slab//
Phys. Rev. 1966. Vol. 150. No. ~2. P. ~ 589 -- 596.

\bibitem {Landau10} {\it Lifshits E. M., Pitaevskii L. P.}
Physical Kinetics,
Butterworth-Heinemann (Jan 1981). P. 625.

\bibitem {S} {\it Sondheimer E. H.}
The mean free path of electrons in metals
//Advances in Physics. 2001. Vol. 50. No. 6. P. ~ 499 - 537.

\bibitem {ly1} {\it Latyshev A. V., Yushkanov A. A.}
 Exact solutions in the theory of the anomalous
 skin -- effect for a plate. Specular boundary conditions.
 Comput. Maths. Math. Phys. 1993. Vol. 33(2), p.p. 229--239.

\bibitem {ly2}
{\it Latyshev A. V., Lesskis A. G., Yushkanov A. A.}
Exact solution to the behavior of the electron plasma in a metal
layer in an alternating electric field,
Theoret. and Math. Phys., 90:2 (1992), 119–126.

\bibitem {Dressel}
{\it Brandt T., H\"{o}vel M., Gompf B., and Dressel M.}
{Temperature - and frequency-dependent optical properties
of ultrathin Au films.}
//Phys. Rev. B. 2008. Vol.78, No. 20. 205409 - 205415. 

\end {thebibliography}
\end{document}